\documentclass[12pt,preprint]{aastex}

\newcommand{\kmsMpc}{{km s$^{-1}$ Mpc$^{-1}$}}

\newcounter{saveeqn}

\shorttitle{ M101 distance}
\shortauthors{Lee and Jang  2012}
\begin{document}

\title{The Distance to M101 Hosting Type Ia SN 2011fe Based on the Tip of the Red Giant Branch } 

\author{Myung Gyoon Lee and In Sung Jang}
\affil{Astronomy Program, Department of Physics and Astronomy, Seoul National University, Gwanak-gu, Seoul 151-742, Korea}
\email{mglee@astro.snu.ac.kr, isjang@astro.snu.ac.kr}


\begin{abstract}
We present a new determination of the distance to M101, host of the type Ia SN 2011fe, based on the tip of the red giant branch method (TRGB). Our determination is based on {\it Hubble Space Telescope} archival $F555W$ and $F814W$ images of nine fields within the galaxy.
Color-magnitude diagrams of arm-free regions in  all fields show a prominent red giant branch (RGB).
We measure the $I$-band magnitudes of the TRGB, obtaining a mean value of   $I_{\rm TRGB}=25.28\pm0.01$ (where the error is a standard error), using an edge-detection method.
We derive a weighted mean value of distance modulus $(m-M)_0=29.30\pm0.01 ({\rm random})\pm0.12 ({\rm systematic})$, corresponding to a linear distance of $7.24\pm0.03\pm0.40 $ Mpc.
While previous estimates for M101 show a large range (TRGB distances of $(m-M)_0=29.05$ to 29.42 and Cepheid distances of  $(m-M)_0=29.04$ to 29.71),  our measurements of the TRGB distances for nine fields show a small dispersion of only 0.02. 
We combine our distance estimate and photometry in the literature to derive absolute peak
magnitudes in optical and near-infrared bands of SN 2011fe.
Absolute maximum magnitudes of SN 2011fe are $\sim0.2$ mag brighter in the optical band and much more in the NIR  than the current calibrations of SNe Ia in the literature.
From the optical maximum magnitudes of SN 2011fe we obtain a value of the Hubble constant, $H_0=65.0\pm0.5({\rm random})\pm5.7({\rm systematic})$ \kmsMpc, slightly smaller than other recent determinations of $H_0$. 

\end{abstract}


\section{Introduction}

M101 (NGC 5457) is a well-known face-on spiral galaxy  (SAB(rs)cd) in the M101 group.
In 2011 a new Type Ia supernova (SN Ia) 2011fe was discovered in this galaxy, being the fourth SN discovered in the same galaxy \citep{nug11,liu12,ric12}.
It is one of the nearest among the known galaxies hosting SNe Ia. SN 2011fe suffers little from interstellar extinction ($A_V\sim 0.04$) and was discovered in less than one day after explosion\citep{nug11}. Therefore it plays an important role for calibrating the absolute luminosity  of SNe Ia as well as for studying the properties of SNe Ia including their progenitors \citep{rei05,nug11,tam12,blo12,rop12,mat12}.


Surprisingly recent measurements of the distance to M101 show a large range, $(m-M)_0=29.04$ to 29.71 (\citet{sha11,mat12,vin12} and references therein). 
Even the measurements based on two primary distance indicators, Cepheids and the tip of the red giant branch (TRGB) \citep{lee93}, show a significant dispersion:
$(m-M)_0=29.04$ to 29.71 for Cepheids \citep{kel96,ken98, ste98, fre01,mac01,sah06,sha11}, 
and $(m-M)_0=29.05$ to 29.42 for the TRGB \citep{sak04,riz07,sha11}.
Note that one of the most recent estimates, \citet{sha11}, shows significant differences in both Cepheid and TRGB distances from previous ones. 
Therefore the distance to M101 is still significantly uncertain in spite of its importance. Consequently, a determination of the Hubble constant based solely on the photometry of SN 2011fe and previously-published distances can yield values ranging from 56 to 76 km s$^{-1}$ Mpc$^{-1}$ \citep{mat12}. 

In this study we present a new determination of the distance to M101 using the TRGB,  from the images for several fields within this galaxy
available in the {\it Hubble Space Telescope} (HST) archive. While previous studies to derive the TRGB distances to M101 were based on one or two fields, we used nine fields. This allows us to estimate the M101 distance with much smaller statistical uncertainty 
than previous studies. 

\section{Data Reduction}

\begin{figure}
\centering
\includegraphics[scale=1.5]{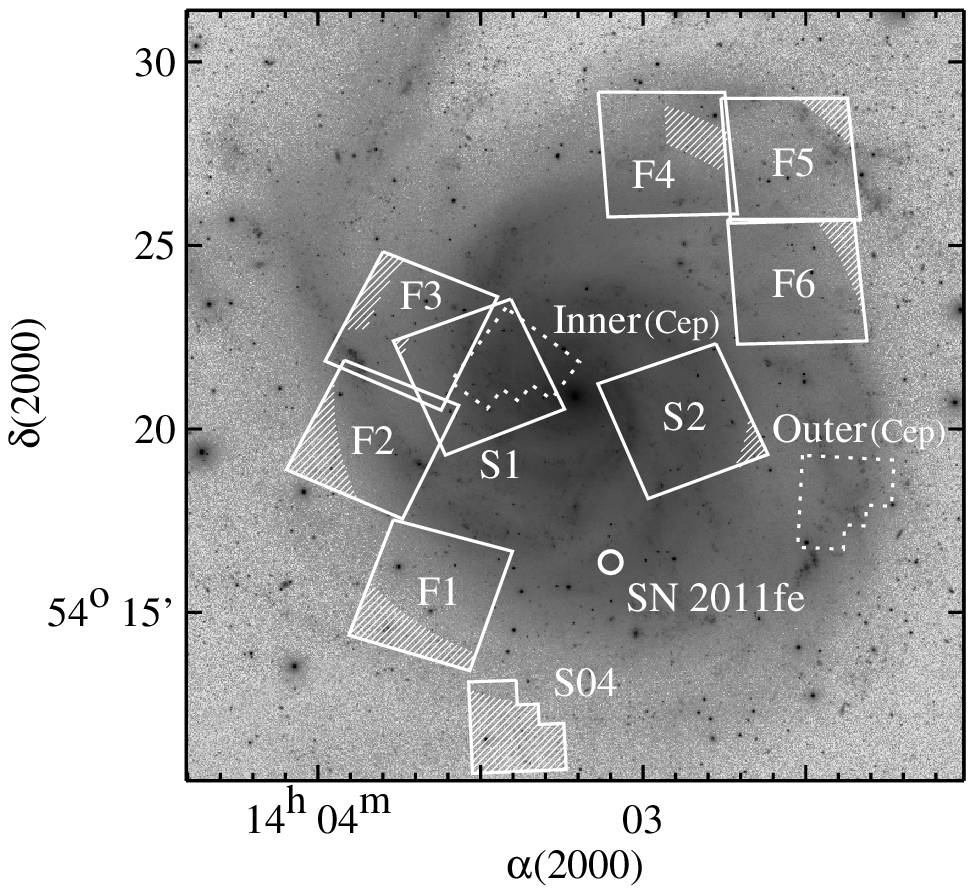} 
\caption{Location of the HST fields marked on the gray scale map of a 21$'$ $\times$ 21$'$ SDSS i-band image of M101. F1, F2, F3, F4, F5, and F6 represent the fields used in this study, S1 and S2 used in \citet{sha11}, and S04 in \citet{sak04}.
Two fields used in previous Cepheid studies \citep{kel96,ken98} 
are labeled by 'outer' and 'inner', respectively.
Only the hatched regions were used in the analysis.
North is up and east to the left. 
}
\label{fig_finder}
\end{figure}

We used  $F555W$ and $F814W$ images of M101 taken with the  HST/Advanced Camera for Surveys (ACS) (Proposal IDs: 9490, 9492, and 10918) and with the WFPC2 (Proposal ID: 8584). 
Figure \ref{fig_finder} displays a finding chart for M101, showing the HST fields used in this study (F1, F2, F3, F4, F5, and F6) and previous studies (S1 and S2 in \citet{sha11} and S04 in \citet{sak04}). 
Note that the S04 field is isolated in the outskirts of the disk, while S1 and S2 are much closer to the center than the other fields.
Two fields used in previous Cepheid studies \citep{kel96,ken98}  
are denoted by 'outer' and 'inner', respectively.
 
We derived instrumental magnitudes of point sources in the ACS images using the IRAF/DAOPHOT package
 that is designed for point spread function (PSF) fitting photometry \citep{ste94}.
We used 2-$\sigma$ as the detection threshold, and derived the PSFs using isolated bright stars in the images. 
We derived aperture corrections using a large aperture
with radius of $1\arcsec$ for several isolated bright stars in the images. 
The uncertainties associated with aperture correction are on average 0.02 mag for both filters.
We transformed the instrumental magnitudes 
into the standard Johnson-Cousins $VI$ magnitudes, 
following equations (1) and (12) for observed magnitudes of \citet{sir05}. 
In the case of WFPC2 images we used Dolphot 
(http://americano.dolphinsim.com/dolphot/).
 The uncertainties associated with the photometric transformations
  are on average 0.02 mag.
 
\section{Results}

\subsection{Color-magnitude Diagrams of Resolved Stars}
 
M101 is an almost face-on galaxy (inclination angle = 17 deg \citep{zar90}) and all HST fields available in the archive are overlapped
with the disk of the galaxy so that they must include a mixture of disk stars and old halo stars. 
To reduce the fraction of young disk stars as much as possible in constructing the color-magnitude diagrams (CMDs) for each field, 
we chose the stars located away from the spiral arms of star-forming regions in each field.
Thus selected regions in each field are marked by the hatched region in Figure 1.  
Foreground reddening toward M101 is known to be very small, $E(B-V)=0.008$ \citep{sch98,sch11}. 
Corresponding values are $A_I=0.013$ and $E(V-I)=0.010$.
Internal reddening for red giants in the arm-free regions of this face-on galaxy is expected to be negligible so that it is not corrected in the  following analysis. 

\begin{figure}
\centering
\includegraphics[scale=0.9]{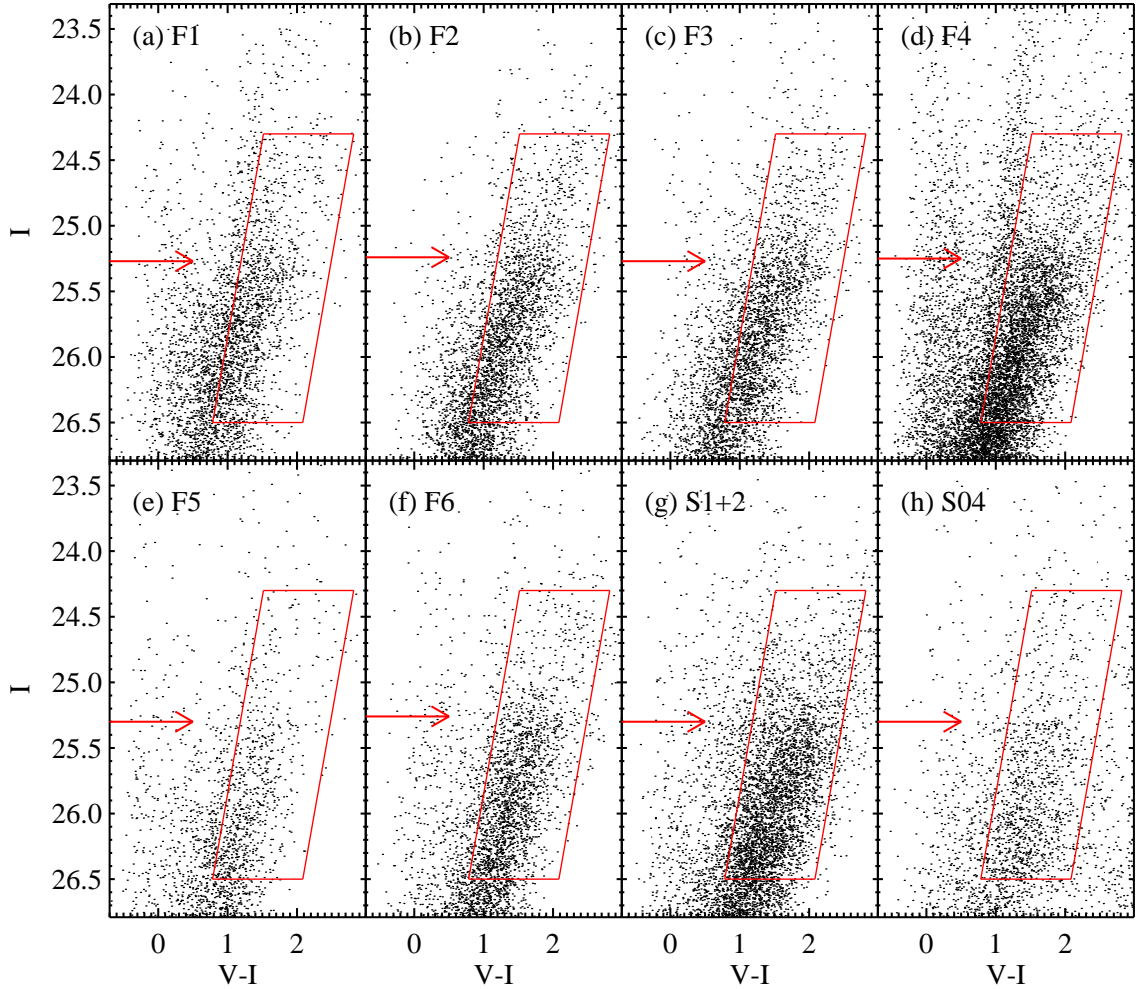} 
\caption{$I-(V-I)$ color-magnitude diagrams of the stars in the arm-free regions in each field of M101. Note the the prominent RGBs are seen as well as other weaker features.
Boxes represent the boundary of the red giants used for deriving
the luminosity functions of the red giant stars in each field. Arrows 
 represent the positions of the TRGB. }
\label{fig_cmd}
\end{figure}

Figure \ref{fig_cmd} displays the CMDs for the stars in the selected regions of F1, F2, F3, F4, F5, F6,  (S1 + S2), and S04 fields. 
The CMDs for all fields show a prominent red giant branch (RGB)
as well as weaker asymptotic giant branch (AGB), massive giants, and blue main sequence. 
These RGBs are useful for estimating the magnitude of the TRGB and the distance to M101. 

\subsection{Distance Estimation}
 
\begin{figure}
\centering
\includegraphics[scale=0.9]{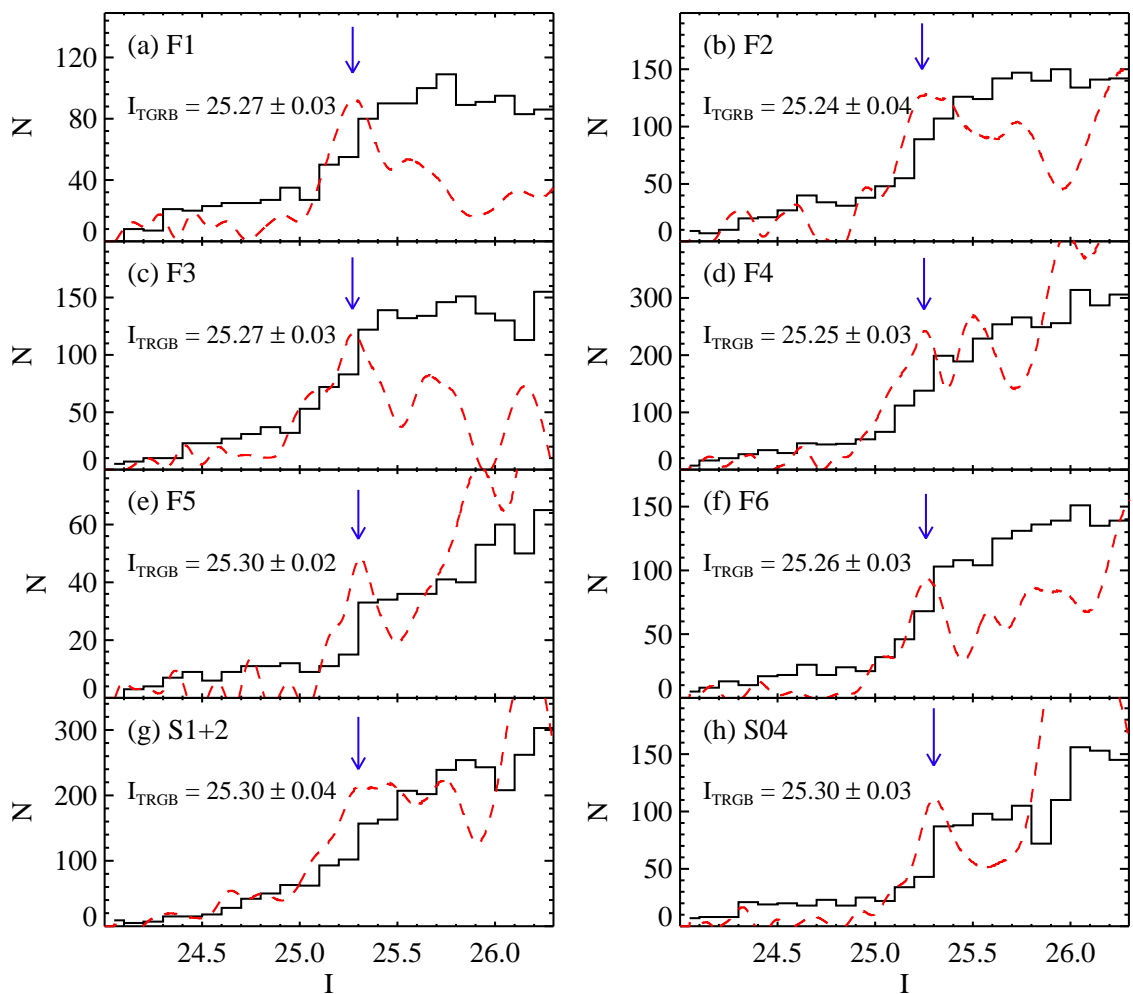} 
\caption{$I$-band luminosity functions of the red giants in each field
of M101 (solid line histograms).  
Dashed lines represent  edge-detection responses, and 
arrows represent the positions of the TRGB. 
}
\label{fig_ilf}
\end{figure}

We determined the distance to the target fields using the TRGB method, which is known to be
an excellent distance indicator for resolved stellar systems \citep{lee93,sak96,men02,bel08,mou10,sal11}. 
First we derived the $I$-band luminosity functions of the red giants in each field  
using the stars inside the boundary marked in Figure \ref{fig_cmd}, and plotted them in Figure \ref{fig_ilf}. 
Figure \ref{fig_ilf} shows that there appears to be a sudden jump at $I \approx 25.3$ mag in each field, 
which corresponds to the TRGB. 

Using the edge-detecting algorithm \citep{sak96,men02,mou10},  we determined the TRGB magnitude more quantitatively.
 We calculated an edge-detection response function 
$E(m)$ ($= \Phi (m + \sigma_m ) -   \Phi (m - \sigma_m )  $ 
where $\Phi (m)$ is the luminosity function of magnitude $m$ and 
$\sigma_m$ is the mean photometric error within a bin of $\pm0.05$ mag about magnitude $m$), 
as shown in Figure \ref{fig_ilf}.
The errors for the TRGB magnitudes were 
determined using bootstrap resampling method with one million simulations. 
In each simulation we resampled randomly
the RGB sample with replacement to make a new sample of the same size. We estimated the TRGB magnitude for
each simulation using the same procedure, and derived the standard deviation of the estimated TRGB magnitudes.
The median color of the TRGB is derived from the colors of the bright red giants close to the TRGB.

%
We used the recent calibration for the absolute magnitude of the TRGB given in \citet{riz07}: 
$M_{{\rm I,TRGB}} = - 4.05(\pm0.02) + 0.217(\pm0.01)( (V-I)_0 -1.6)$ (where $(V-I)_0$ is a reddening corrected
color of the TRGB), 
which  is very similar to that given in \citet{tam08}. 
Then we calculate the distance modulus using $(m-M)_0 = I_{\rm 0, TRGB} - M_{\rm I, TRGB}$.

We also derived the distance using the composite magnitude $T$ for the RGB introduced in \citet{mad09}:
$T = I_{0, TRGB} -  0.20 [(V-I)_{0} -1.5)]$ 
and $(m-M)_0 = T -  M_{I,TRGB}$ $=  I_{0, TRGB} - 0.20 (V-I)_{0}  +4.35$. 
This calibration is based on the absolute TRGB magnitude of $M_{I,TRGB} = - 4.05$ 
for $(V-I)_0=1.5$, 
similar to the calibration in \citet{riz07}.

Table 1 lists a summary of the distance determination
for the fields in M101 derived in this study, which is also shown in Figure 4.
Table 1 includes $I$-band magnitudes of the TRGB,  $T$ magnitudes of the RGB, 
$(V-I)$ color of the RGB (before foreground reddening correction), absolute $I$-band magnitudes of the TRGB, and
distance moduli from $I$-band magnitudes and $T$ magnitudes.
Remarkably the $I$-band magnitudes of the TRGB derived for all nine fields show a small range from 25.24 (F2) to 25.30 (F5, S1, S2 and S04). 
A weighted mean value is derived to be   $I_{\rm TRGB}=25.28\pm0.01$ (where the error is a standard error) 
with a standard deviation of only 0.02. 
Similarly the distance moduli for all fields show a small range from  $(m-M)_0=29.26\pm0.03$ (F4) to $29.33\pm0.02$ (F5). 
We derive a weighted mean value of distance modulus $(m-M)_0=29.30\pm0.01$  where 0.01 is a random error, 
corresponding to a linear distance of $7.24\pm0.03 $ Mpc.  
Its systematic error is estimated to be 0.12, considering (a) the TRGB calibration error of 0.12 \citep{bel01, bel04,mag08},
(b) the aperture correction error of 0.02, (c) and the standard calibration error of 0.02 for ACS/WFC \citep{sir05} and 0.07 for WFPC2 \citep{ste98a}.
Similar results are obtained from  the $T$ magnitude method, $(m-M)_0=29.26\pm0.01$,
but with three times larger standard deviation (0.07).
We adopt the results from the traditional $I$-band magnitude method,  as the distance to M101, 
which shows a smaller scatter than the $T$ magnitude results.

\section{Discussion}


\subsection{Comparison with Previous Distance Estimates}

\begin{figure}
\centering
\includegraphics[scale=1.1]{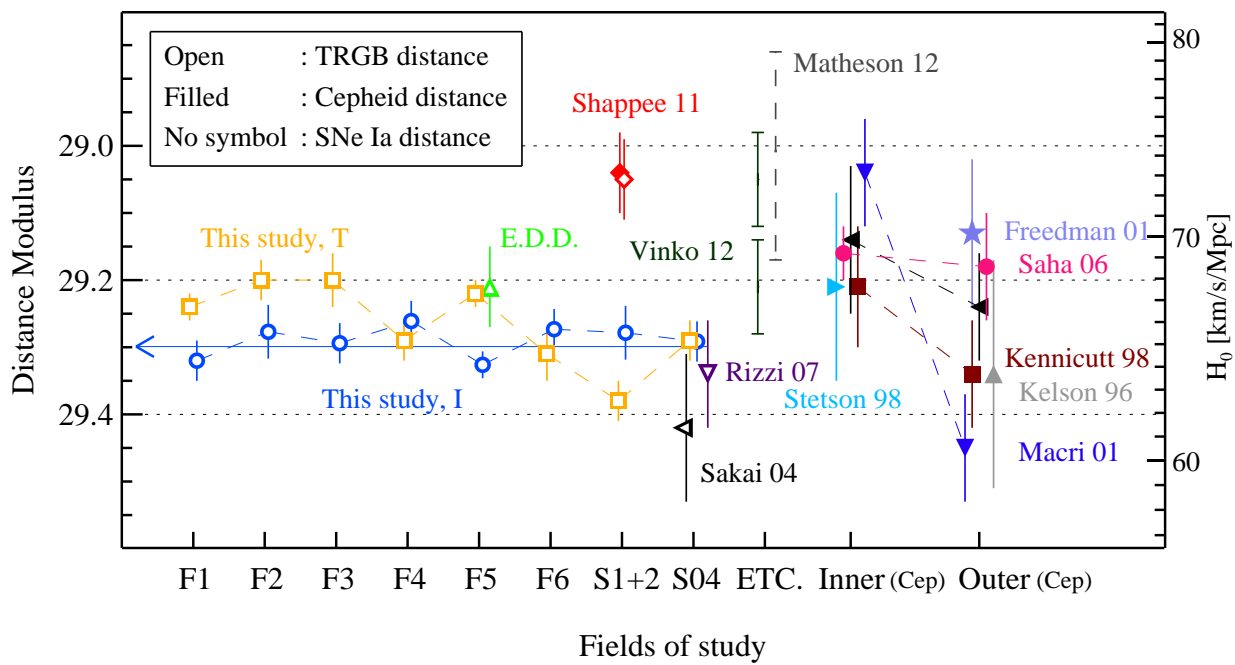} 
\caption{A comparison of the distance measurements for M101
based on the TRGB (open symbols), Cepheids (filled symbols), and SN 2011fe (bars without any symbol) versus fields of study:
F1-F6 in this study, S1 and 2 in \citet{sha11}, S04 in \citet{sak04}, inner and outer fields for Cepheids, and others. The long horizontal arrow represents the weighted mean value of the TRGB distance estimates derived in this study. The bar at the right end represents the $H_0$ value depending on the M101 distance, based on the relation in \citet{rei05}.} 
\label{fig_comp}
\end{figure}

We compare our estimates for the distance to M101 with those in the literature based on TRGB, Cepheids, and SN Ia in Figure \ref{fig_comp}.
Recently \citet{sha11} derived a TRGB distance from the analysis of S1 and S2 fields, 
$(m-M)_0=29.05 \pm0.06 ({\rm random}) \pm0.12 ({\rm systematic})$. 
This value is 0.3 to 0.4 mag smaller than the previous TRGB distance estimates by \citet{sak04} and \citet{riz07}.
The TRGB magnitudes for M101 derived in the previous studies are
$I_{0,\rm TRGB} = 25.40\pm0.04$ in \citet{sak04},  
$25.29\pm0.08$ in \citet{riz07}, and $T = 25.00\pm0.06$ in \citet{sha11}.
Therefore the large difference among the previous TRGB distance estimates are mainly due to the difference in the measured TRGB magnitudes. 
Our mean value of the TRGB magnitudes (after foreground extinction correction), $I_{0,\rm TRGB}=25.27\pm0.01$ (and $T_{\rm RGB}=25.33\pm0.03$ for (S1+S2) fields) is close to the value in \citet{riz07},
0.13 mag brighter than the value in \citet{sak04}, and 0.33 mag fainter 
than the value in \citet{sha11}. 
Our values for the S04 field ($I_{0,\rm TRGB}=25.29\pm0.03$) and for the sum of S1 and S2 fields ($I_{0,\rm TRGB}=25.29\pm0.04$) are similar to the mean value of nine fields.
The large difference in the TRGB magnitudes between this study and \citet{sha11} 
is probably due to contamination by disk stars in the sample used by the latter.

The Cepheid distance estimates for M101 derived in the previous studies show a large range. 
Two HST fields in M101 (inner field and outer field, as shown Figure 1) were used for Cepheids 
in the previous studies. They  show on average $\sim 0.2$ smaller values for the inner field \citep{ste98,sah06, ken98, mac01, sha11}
than those for the outer field \citep{kel96, ken98, mac01, sah06}.
This difference between the inner and outer  fields
may be considered to be mainly due to metallicity gradient in the galaxy disk. 
However, the distance modulus difference shows a large range from almost zero \citep{sah06}
to ∼ 0.4 mag \citep{mac01}, 
depending on the authors. This difference has been explained. \citet{sah06} adopted a metallicity correction very similar to \citet{fre01}, 
which led them to Cepheid distances of 29.16 for the inner field and 29.18 for the outer field. These compare well to the $H_0$ Key Project final value of $29.13\pm0.11$ mag \citep{fre01}. 
\citet{mac01} carried out artificial star tests and showed that their NICMOS photometry for the inner field was by $\sim$0.2 mag affected by blending.
Cepheid distance estimates show a large scatter (of ∼ 0.3 mag) even for the outer field where blending effect is less severe, depending on the authors. 
The distance estimates by \citet{fre01,sak04,sah06} 
show a good agreement among them, but they are $\sim$0.2 mag shorter than those by \citet{kel96,ken98,mac01}. 
 The cause for this scatter needs to be investigated.

In contrast, our measurements of the TRGB magnitude for nine fields show a remarkably small dispersion. 
Note that our measurements show an excellent agreement among independent photometry
(for example, between DAOPHOT photometry of ACS images and DOLPHOT photometry of WFPC2 images, among three different sets of HST images (F fields, S1 and S2 fields, and S04 fields), and among all nine fields).

\subsection{The Absolute Calibration of SNe Ia and the Hubble Constant}

We can use the distance measurement for M101 to check the calibration of SN Ia and to derive
a value for the Hubble constant.
Using two SN Ia optical light curve fitting methods with $BVRI$ photometry of SN 2011fe,
 \citet{vin12} obtained  two estimates for the distance to M101 (assuming $H_0 = 73$ km s$^{-1}$ Mpc $^{-1}$): 
$(m-M)_0 = 29.21\pm0.07$ from the MLCS2k2 method \citep{jha07} and
 $(m-M)_0 = 29.05\pm0.08$ from the SALT2 method \citep{guy07}. 
 They concluded that the difference between these two estimates  is considered to be due to difference in the zero point calibration of the fiducial SN peak magnitude in the two methods. 
Our result is closer to the value from the  MLCS2k2 method \citep{jha07},  indicating that the calibration of the MLCS2k2 method is closer to the TRGB calibration, for the adopted $H_0 = 73$ km s$^{-1}$ Mpc $^{-1}$.
 
\citet{ric12} presented  optical maximum magnitudes of SN 2011fe: 
 $B_{\rm max} = 10.00\pm0.02$, $V_{\rm max} = 9.99\pm0.01$, $R_{\rm max} = 9.99\pm0.02$ and $I_{\rm max} = 10.21 \pm0.03$,
and derived absolute magnitudes of SN 2011fe
adopting a distance modulus of $(m-M)_0 = 29.10\pm0.15$.
They derived also similar values considering the decline rate of $\Delta m_{15} (B) = 1.21\pm0.03$.
%
Adopting our distance measurement, 
we derive absolute magnitudes of SN 2011fe that are $\sim$0.2 mag brighter than those in \citet{ric12}: 
$M_{B, {\rm max}} = -19.41\pm0.13$, $M_{V, {\rm max}} = -19.39\pm0.13$, and $M_{R, {\rm max}} = -19.38\pm0.13$, and $M_{I, {\rm max}} = -19.14\pm0.13$. 
($M_{B, {\rm max}} = -19.45\pm0.13$, $M_{V, {\rm max}} = -19.38\pm0.13$, and $M_{R, {\rm max}} = -19.39\pm0.13$, and
$M_{I, {\rm max}} = -19.12\pm0.13$ after  the decline rate correction).  
%
The derived value of $M_{V, {\rm max}} = -19.38 \pm 0.13$  
is between of the values derived from nearby SNe Ia in the previous studies:   
recent values in \citet{rie11} ($M_{V,{\rm max}} = -19.15\pm0.07$) and  old values in \citet{gib00} (
$M_{V,{\rm max}}=-19.46\pm0.05$)
that are similar to those in \citet{san06}.
%

Near-infrared (NIR)  photometry of SN Ia is potentially a very promising tool for cosmology \citep{kat12,bar12}. 
However, recent calibrations of the NIR absolute magnitudes of SN Ia show large 
 differences ( $\sim$ 0.4 mag) \citep{woo08,fol10,bur11,man09, man11,kat12}, as summarized in \citet{mat12}.
Our distance measurement for M101 is very useful for calibrating these NIR zero points of the SN Ia templates.
 \citet{mat12} derived a large range of distance estimates from  $JHK_S$ photometry of SN 2011fe:  
 $(m-M)_0 = 28.86$ to 29.17, depending on the adopted calibration. These values are much smaller than our distance measurement.

\citet{mat12} presented $JHK_s$ maximum magnitudes in each band for SN 2011fe:
$J_{\rm max} = 10.51\pm0.04$, $H_{\rm max}  = 10.75\pm0.04$, and $K_{s,\rm max} = 10.64\pm0.04$,  and
$JHK_s$ magnitudes at the $B$-band maximum time for SN 2011fe:
$J_{B_{\rm max}} = 10.62\pm0.04$, $H_{B_{\rm max}} = 10.85\pm0.04$, and $K_{s,B_{\rm max}}  = 10.68\pm0.05$.
($A_K=0.01$ is ignored here, as in \citet{mat12} (see their Fig. 3)).

From these values with our distance estimate we derive the NIR absolute magnitudes for SN 2011fe:
$M_{J,{\rm max}} = -18.79\pm0.14$, $M_{H,{\rm max}} = -18.55\pm0.14$, and $M_{K_s, {\rm max}} = -18.66 \pm0.13$
($M_{J,{B_{\rm max}}} = -18.68\pm0.14$, $M_{H,{B_{\rm max}}} = -18.45\pm0.14$, and $M_{K_s, {B_{\rm max}}} = -18.62 \pm0.14$ for the $B$-band maximum time).
These values are $\sim$0.2 mag or more brighter than recent calibration of the NIR magnitudes for SN Ia available in the literature \citep{woo08,fol10,bur11,man09,man11,kat12}.

 If we use 
 absolute magnitudes of SN 2011fe derived in this study,  
 we obtain a value for the Hubble constant using the equation
 log $H_0 = 0.2 M_\lambda ({\rm max}) + C_\lambda +5$ 
 (where $C_B=0.693\pm0.004$,  $C_V=0.688\pm0.004$, and $C_I=0.637\pm0.004$) \citep{rei05}, 
 $H_0 = 65.0 \pm0.5{\rm (random)}  \pm5.7 {\rm (systematic)}$ \kmsMpc 
 (we included the internal luminosity dispersion of SNe Ia of 0.14 mag (\citet{tam12}) for calculating the systematic error). 
 This value is   similar to    that given in \citet{tam12} based on six SNe Ia including SN 2011fe, $H_0 = 64.0 \pm1.6 \pm2.0$  \kmsMpc, 
 but smaller than other recent determinations \citep{rie11,fre12}, 
 $H_0 = 74 \pm3 $ km s$^{-1}$ Mpc$^{-1}$.
This implies three possibilities: (a) that the optical maximum magnitudes of SN 2011fe may be $\sim0.2$ mag brighter than typical SNe Ia, (b) that the recent calibration of optical maximum magnitudes of SN Ia may be $\sim$0.2 mag fainter, or  (c) that the value  for the Hubble constant may be somewhat  lower than the values derived recently in other studies.

\section{Summary}
We present a new determination of the distance to M101, host of the Type Ia SN 2011fe, using the tip of the red giant branch method (TRGB) from $F555W$ and $F814W$ images of nine fields within this galaxy in the Hubble Space Telescope archive.
Primary results are as follows.

\begin{itemize}
\item Color-magnitude diagrams of arm-free regions in  all fields show a prominent RGB.

\item We measured the $I$-band magnitudes of the TRGB, obtaining a mean value of   $I_{\rm TRGB}=25.28\pm0.01$  with a standard deviation of only 0.02, using an edge-detection method.
From this we derive a weighted mean value of distance modulus $(m-M)_0=29.30\pm0.01({\rm random})$,
 corresponding to a linear distance of $7.24\pm0.03 $ Mpc.
Its systematic error is estimated to be 0.12.
Our measurements of the TRGB distances for nine fields show a small dispersion of only 0.02, much smaller than the previous estimates.

\item With this distance value we derive the optical and NIR absolute maximum magnitudes of SN 2011fe (at the maximum of each band and at the $B$-band maximum time).
Absolute magnitudes of SN 2011fe are $\sim$0.2 mag brighter in the optical band and much more in the NIR  than the recent calibrations of SN Ia in the literature.
 
\item 
From the optical magnitudes of SN 2011fe and our distance measurement for M101 we obtain a value of the Hubble constant, $H_0=65.0 \pm0.5({\rm random}) \pm5.7({\rm systematic})$ \kmsMpc, somewhat smaller than other recent determinations \citet{rie11,fre12}.
This implies  (a) that optical maximum magnitudes of SN 2011fe may be $\sim$0.2 mag brighter than typical SNe Ia, (b) that the recent calibration of optical maximum magnitudes of SN Ia may be $\sim0.2$ mag fainter, or
 (c) that the value  for the Hubble constant may be somewhat  lower than other recent determinations.
 
\end{itemize}

\bigskip

The authors are grateful to the anonymous referee for his/her comments that improved the original manuscript.
This is supported in part by the Mid-career Researcher Program through an NRF grant funded by the MEST (No.2010-0013875).



\clearpage

\begin{deluxetable}{rccccccccc}
\tabletypesize{\scriptsize}
\tabletypesize{\footnotesize} 
\setlength{\tabcolsep}{0.05in}
\rotate
\tablecaption{Summary of TRGB distance measurements of M101}
\tablewidth{0pt}

\tablehead{ \colhead{Field} & \colhead{Region} & \colhead{ $I_{TRGB}$ } & \colhead{ $T_{RGB}$}  & \colhead{ $(V-I)_{TRGB}$} & \colhead{ $M_{I,TRGB}$} & \colhead{ $(m-M)_{0,I}$} & \colhead{ $(m-M)_{0,T}$}}

\startdata

F1 & $R_{GC}\ge7'.5$        & $25.27\pm0.03$ & $25.19\pm0.02$ & $1.54\pm0.04$ & $-4.065\pm0.009$ & $29.32\pm0.03$ & $29.24\pm0.02$ \\
F2 & $R_{GC}\ge6'.5$        & $25.24\pm0.04$ & $25.15\pm0.03$ & $1.60\pm0.05$ & $-4.052\pm0.011$ & $29.28\pm0.04$ & $29.20\pm0.03$ \\
F3 & $R_{GC}\ge5'.9$, Dec $\ge 54\,^{\circ}.4   $  & $25.27\pm0.03$ & $25.15\pm0.04$ & $1.66\pm0.05$ & $-4.039\pm0.011$ & $29.30\pm0.03$ & $29.20\pm0.04$ \\
F4 & $7'.3\le R_{GC}<8'.3$, R.A. $< 210\,^{\circ}.73$ & $25.25\pm0.03$ & $25.24\pm0.03$ & $1.72\pm0.06$ & $-4.026\pm0.013$ & $29.26\pm0.03$ & $29.29\pm0.03$ \\
F5 & $R_{GC}\ge10'.1$      & $25.30\pm0.02$ & $25.17\pm0.02$ & $1.65\pm0.05$ & $-4.041\pm0.011$ & $29.33\pm0.02$ & $29.22\pm0.02$ \\
F6 & $R_{GC}\ge8'.1$        & $25.26\pm0.03$ & $25.26\pm0.04$ & $1.71\pm0.05$ & $-4.028\pm0.011$ & $29.28\pm0.03$ & $29.31\pm0.04$ \\
S1 & $R_{GC}\ge4'.75$      & $25.30\pm0.04$ & $25.29\pm0.03$ & $1.81\pm0.04$ & $-4.007\pm0.009$ & $29.29\pm0.04$ & $29.34\pm0.03$ \\
S2 & $R_{GC}\ge4'.75$      & $25.30\pm0.04$ & $25.33\pm0.03$ & $1.91\pm0.03$ & $-3.985\pm0.007$ & $29.27\pm0.04$ & $29.38\pm0.03$ \\
S1+2 & $R_{GC}\ge4'.75$   & $25.30\pm0.04$ & $25.33\pm0.03$ & $1.87\pm0.05$ & $-3.994\pm0.011$ & $29.28\pm0.04$ & $29.38\pm0.03$ \\
S04 & $R_{GC}\ge8'.5$      & $25.30\pm0.03$ & $25.24\pm0.03$ & $1.81\pm0.05$ & $-4.007\pm0.011$ & $29.29\pm0.03$ & $29.29\pm0.03$ \\
\multicolumn{2}{c}{Weighted mean of F1$\sim$6, S1+2 and S04} & $25.28\pm0.01$ & $25.21\pm0.01$& $1.68\pm0.02$ & $-4.034\pm0.004$ & $29.30\pm0.01$ & $29.26\pm0.01$ \\
\enddata
\end{deluxetable}

\clearpage

%
%
%
%


\end{document}